\pdfoutput=1
\RequirePackage{ifpdf}
\ifpdf % We~are running pdfTeX in pdf mode
\documentclass[pdftex]{sigma}
\else
\documentclass{sigma}
\fi

\def\Xint#1{\mathchoice
{\XXint\displaystyle\textstyle{#1}}%
{\XXint\textstyle\scriptstyle{#1}}%
{\XXint\scriptstyle\scriptscriptstyle{#1}}%
{\XXint\scriptscriptstyle\scriptscriptstyle{#1}}%
\!\int}
\def\XXint#1#2#3{{\setbox0=\hbox{$#1{#2#3}{\int}$}
\vcenter{\hbox{$#2#3$}}\kern-.5\wd0}}

\def\dashint{\Xint-}

\numberwithin{equation}{section}

\newtheorem{Theorem}{Theorem}[section]
\newtheorem{Lemma}[Theorem]{Lemma}
 { \theoremstyle{definition}
\newtheorem{Remark}[Theorem]{Remark} }

\begin{document}
\allowdisplaybreaks

\newcommand{\arXivNumber}{1809.07362}

\renewcommand{\PaperNumber}{139}

\FirstPageHeading

\ShortArticleName{Exact Formulas of the Transition Probabilities}

\ArticleName{Exact Formulas of the Transition Probabilities\\ of the Multi-Species Asymmetric Simple Exclusion\\ Process}

\Author{Eunghyun LEE}

\AuthorNameForHeading{E.~Lee}

\Address{Nazarbayev University, Nur-sultan, Kazakhstan}
\Email{\href{mailto:eunghyun.lee@nu.edu.kz}{eunghyun.lee@nu.edu.kz}}
\URLaddress{\url{https://sites.google.com/a/nu.edu.kz/eunghyun-lee-s-homepage/}}

\ArticleDates{Received September 17, 2020, in final form December 15, 2020; Published online December 20, 2020}

\Abstract{We find the formulas of the transition probabilities of the $N$-particle multi-species asymmetric simple exclusion processes (ASEP), and show that the transition probabilities are written as a determinant when the order of particles in the final state is the same as the order of particles in the initial state.}

\Keywords{ASEP; multi-species ASEP; integrable probability}

\Classification{82C22; 60J27}

\section{Introduction}
The exact formulas of the transition probabilities of the \textit{exactly solvable} models may be a good starting point to study interesting distributions and their asymptotic behaviours \cite{Korhonen-Lee-2014,Lee-2010,Lee-2012,Lee-Wang-2018, Schutz-1997,Tracy-Widom-2008,Wang-Waugh-2016}. In the multi-species versions of these models, particles in the system may belong to different classes and there is a hierarchy of these classes. In this paper, we consider the multi-species version of the asymmetric simple exclusion processes (ASEP). (See \cite{Kuan-2018,Kuniba-Mangazeev-Maruyama-Okado-2016} for the multi-species version of the totally asymmetric zero range processes.) We consider $N$-particle systems which may consist of up to $N$ species, labelled $1,\dots, N$. The rules for the multi-species ASEP are as follows: a particle at $x \in \mathbb{Z}$ waits an exponential random time with rate $1$ and then chooses $x+1$ with probability $p$ or $x-1$ with probability $q=1-p$ to jump. If the particle at $x$ belongs to species $l$ and the chosen site to jump is already occupied by another particle belonging to species $l'\geq l$, then the jump is prohibited, but if $l'<l$, the particle belonging to $l$ jumps to the chosen site by interchanging sites with the particle belonging to $l'$. According to this rule, we may view an empty site as a ``\textit{particle}" labelled $0$.

A state of the process is denoted by a pair $(X, \pi)$ where $X = (x_1,\dots, x_N) \in \mathbb{Z}^N$ with $x_1<\cdots <x_N$ and $\pi= \pi_1\pi_2~\cdots~\pi_N$ is a permutation of a multi-set $\mathcal{M} = [i_1,\dots,i_N]$ with elements taken from $\{1,\dots, N\}$. Each $x_i$ represents the position of the $i$th leftmost particle and~$\pi_i$ represents the species the $i$th leftmost particle belongs to.

The purposes of this paper are to provide the explicit formula of the probability $P_{(Y,\nu)}(X,\pi;t)$ that the system is in $(X,\pi)$ at time $t$, given an initial state $(Y,\nu)$ for each $\mathcal{M}$, and to show that the formula of $P_{(Y,\nu)}(X,\pi;t)$ is in the form of a determinant in the case that all particles move only in one direction and $\nu = \pi$. If $\mathcal{M} = [i,i,\dots,i]$, then the permutation of $[i,i,\dots,i]$ is uniquely $ii\,\cdots\,i$, so the system is the ASEP. If $\mathcal{M} = [1,2,\dots,N]$, then all $N$ particles in the system belong to all different species. The exact formulas of the transition probabilities of the ASEP, written $P_Y(X;t)$, were obtained by Tracy and Widom~\cite{Tracy-Widom-2008}, generalizing Sch\"{u}tz's formulas~\cite{Schutz-1997} for the totally asymmetric simple exclusion process (TASEP). Tracy and Widom's formula for the transition probabilities of the ASEP is an $N$-fold contour integral
\begin{equation}\label{103367}
P_Y(X;t)= \dashint_c \cdots \dashint_c \sum_{\sigma \in {S}_N}A_{\sigma}\prod_{i=1}^N\big(\xi_{\sigma(i)}^{x_i - y_{\sigma(i)}-1}{\rm e}^{\varepsilon(\xi_i)t}\big) {\rm d}\xi_1 \cdots {\rm d}\xi_N,
\end{equation}
where the notation $\dashint_c$ implies $\frac{1}{2\pi {\rm i}}\int_c$ and the contour $c$ is a circle centered at the origin with \textit{sufficiently small} radius. The sum inside the integrals in~(\ref{103367}) is over all permutations~$\sigma$ in the symmetric group $S_N$ and
\begin{equation}\label{104567}
A_{\sigma} = \prod_{(\beta,\alpha)} S_{\beta\alpha},
\end{equation}
where
\begin{equation*}
S_{\beta\alpha} = -\frac{p+q\xi_{\alpha}\xi_{\beta} - \xi_{\beta}}{p+q\xi_{\alpha}\xi_{\beta} - \xi_{\alpha}}
\end{equation*}
and
\begin{equation*}
\varepsilon(\xi_i) = \frac{p}{\xi_i} + q\xi_i- 1.
\end{equation*}
The product in (\ref{104567}) is over all inversions $(\beta,\alpha)$ with $\beta >\alpha$ in $\sigma$, and if $\sigma$ is the identity permutation, written ${\rm Id}$, then we define $A_{\sigma} = 1$.
Tracy and Widom have shown that there is a~formula analogous to~(\ref{103367}) for the transition probabilities of the multi-species ASEP, written as
\begin{gather*}%\label{105067}
P_{(Y,\nu)}(X,\pi;t)= \dashint_c\cdots\dashint_c \sum_{\sigma \in {S}_n}A_{\sigma}^{\nu,\pi}\prod_{i=1}^N\big(\xi_{\sigma(i)}^{x_i - y_{\sigma(i)}-1}{\rm e}^{\varepsilon(\xi_i)t}\big) {\rm d}\xi_1 \cdots {\rm d}\xi_N,
\end{gather*}
in our notation \cite{Tracy-Widom-2013}. But, the formula of $A_{\sigma}^{\nu,\pi}$ was not given explicitly except a special case in \cite[p.~458]{Tracy-Widom-2013}. In this paper, we provide a method to find the formula of $A_{\sigma}^{\nu,\pi}$. This is elaborated in Section~\ref{prelim}. In Section \ref{det}, we find a determinantal expression for transition probabilities.

\section{Transition probabilities}\label{prelim}
We first review the basic concepts used in \cite{Chatterjee-Schutz-2010,Lee-2017,Lee-2018} to extend the results for the TASEP with second class particles there to the multi-species ASEP.
The state space of the multi-species ASEP with $N$ particles is countable, so we may view $P_{(Y,\nu)}(X,\pi;t)$ as a matrix element of an infinite matrix, denoted $\mathbf{P}(t)$, which is a member of a probability semigroup $\{\mathbf{P}(t)\colon t \geq 0\}$. We assume that the rows are labelled by $(X,\pi)$ and the columns are labelled by $(Y,\nu)$, following a~rule that if $\pi\prec \pi'$ where $\prec$ is the lexicographical order, then the row $(X,\pi')$ is below the row $(X,\pi)$ for a given $X$, and similarly, the column $(Y,\nu')$ is to the right of the column $(Y,\nu)$ if $\nu \prec\nu'$ for a given $Y$. For fixed $X$ and $Y$, let $\mathbf{P}_Y(X;t)$ be a sub-matrix of~$\mathbf{P}(t)$, consisting of the rows labelled $(X, \cdot)$ and the columns labelled $(Y,\cdot)$ of $\mathbf{P}(t)$. Then, $ \mathbf{P}_Y(X;t)$ is the $N^N \times N^N$ matrix in the form of
\begin{gather*} %\label{456pm618}
 \mathbf{P}_Y(X;t) =
 \left[\!
 \begin{matrix}
 P_{(Y,1\cdots1)}(X,1\cdots11;t)\!\! & P_{(Y,1\cdots12)}(X,1\cdots11;t)\!\! & \!\!\cdots\!\! & P_{(Y,N\cdots N)}(X,1\cdots11;t) \\[2pt]
 P_{(Y,1\cdots1)}(X,1\cdots12;t)\!\! & P_{(Y,1\cdots12)}(X,1\cdots12;t)\!\! & \!\!\cdots\!\! & P_{(Y,N\cdots N)}(X,1\cdots12;t) \\[2pt]
 \vdots & \vdots & \!\!\ddots\!\! & \vdots \\[2pt]
 P_{(Y,1\cdots1)}(X,N\cdots N;t)\!\! & P_{(Y,1\cdots12)}(X,N\cdots N;t)\!\! & & P_{(Y,N\cdots N)}(X,N\cdots N;t)
 \end{matrix}\!
 \right]\!.
\end{gather*}
The labels of the rows and the columns of this matrix are the permutations of the multi-sets of cardinality~$N$ with elements taken from $\{1,\dots,N\}$ in the lexicographical order from the top to the bottom and from the left to the right, respectively.

 The semigroup $\{\mathbf{P}(t)\colon t \geq 0\}$ is uniform, and $\mathbf{P}(t)$ is the unique solution to the forward equation which is an element-wise matrix differential equation
\begin{equation*}%\label{335pm620}
\frac{{\rm d}}{{\rm d}t}\mathbf{P}(t) = \mathbf{G}\mathbf{P}(t)
\end{equation*}
 subject to the initial condition $\mathbf{P}(0) = \mathbf{I}_{\infty}$ where $\mathbf{I}_{\infty}$ is the infinite identity matrix and $\mathbf{G}$ is the generator. (We use the notation $\mathbf{I}_n$ for $n \times n$ identity matrix and $\mathbf{0}_n$ for $n \times n$ zero matrix.) The initial condition $\mathbf{P}(0) = \mathbf{I}_{\infty}$ implies that the sub-matrices $\mathbf{P}_Y(X;t)$ satisfy
 \begin{equation*}
 \mathbf{P}_Y(X;0) = \begin{cases}
 \mathbf{I}_{N^N} & \textrm{if}~X=Y, \\
 \mathbf{0}_{N^N} & \textrm{otherwise}.
 \end{cases}%\label{749am724a}
 \end{equation*}
By the simple extension of Section~2.1 in \cite{Lee-2018} for the TASEP with second class particles to the ASEP with second class particles, we can obtain a matrix which generalizes~(2.11) in~\cite{Lee-2018},
\begin{equation*}%\label{123am621}
\widetilde{\mathbf{R}}_{\beta\alpha} = \left[
 \begin{matrix}
 S_{\beta\alpha} & 0 & 0 & 0 \\
 0 & P_{\beta\alpha} & pT_{\beta\alpha} & 0 \\
 0 & qT_{\beta\alpha} & Q_{\beta\alpha} & 0 \\
 0 & 0 & 0 &S_{\beta\alpha}
 \end{matrix}
 \right],
\end{equation*}
where
\begin{alignat*}{3}
& S_{\beta\alpha} = -\frac{p+q\xi_{\alpha}\xi_{\beta} - \xi_{\beta}}{p+q\xi_{\alpha}\xi_{\beta} - \xi_{\alpha}},\qquad && P_{\beta\alpha} = \frac{(p-q\xi_{\alpha})(\xi_{\beta}-1)}{p+q\xi_{\alpha}\xi_{\beta} - \xi_{\alpha}}, & \nonumber\\
& T_{\beta\alpha} = \frac{\xi_{\beta}-\xi_{\alpha}}{p+q\xi_{\alpha}\xi_{\beta} - \xi_{\alpha}},\qquad && Q_{\beta\alpha} =\frac{(p-q\xi_{\beta})(\xi_{\alpha}-1)}{p+q\xi_{\alpha}\xi_{\beta} - \xi_{\alpha}}.&%\label{1100am911}
\end{alignat*}
 Notice that the denominators of the nonzero elements in $\widetilde{\mathbf{R}}_{\beta\alpha}$ have the same form as the denominator of $S_{\beta\alpha}$ in \cite{Tracy-Widom-2008,Tracy-Widom-2011}. This fact makes it easy to prove \textit{the initial condition} later. The matrix $\widetilde{\mathbf{R}}_{\beta\alpha}$ is a building block for $N$-particle system. In $N$-particle ASEP with multi-species, a matrix corresponding to ${\mathbf{B}}$ in~(2.6) of~\cite{Lee-2018} which is for a two-particle TASEP is an $N^2 \times N^2$ matrix~${\mathbf{B}}$ with
\begin{equation*}%\label{336pm621}
\big[\mathbf{B}\big]_{ij,kl} = \begin{cases}
1& \textrm{if}~ij=kl~\textrm{with}~i=j,\\
p& \textrm{if}~\textrm{either}~ij=kl~\textrm{or}~ij=lk~\textrm{with}~i<j,\\
q& \textrm{if}~\textrm{either}~ij=kl~\textrm{or}~ij=lk~\textrm{with}~i>j,\\
0 &\textrm{for all other cases},
\end{cases}
\end{equation*}
where $ij$ and $kl$ are labels for rows and columns, represented by $11,12,\dots, NN$. In the same way as $\widetilde{\mathbf{R}}_{\beta\alpha}$ is obtained, we define an $N^2 \times N^2$ matrix $\mathbf{R}_{\beta\alpha}$ by
\begin{equation*}%\label{300pm622}
\mathbf{R}_{\beta\alpha}= -\big[(p+q\xi_{\alpha}\xi_{\beta})\mathbf{I}_{N^2} - \xi_{\alpha}\mathbf{B}\big]^{-1}\big[(p+q\xi_{\alpha}\xi_{\beta})\mathbf{I}_{N^2} - \xi_{\beta}\mathbf{B}\big],
\end{equation*}
where $\alpha, \beta = 1,\dots, N$ and $\alpha \neq \beta$. Here, $\mathbf{I}_{N^2}$ is the $N^2 \times N^2$ identity matrix.
The entries of~$\mathbf{R}_{\beta\alpha}$ are given by
\begin{equation*}%\label{625pm724}
\big[\mathbf{R}_{\beta\alpha}\big]_{ij,kl} = \begin{cases}
S_{\beta\alpha}& \textrm{if}~~ij=kl~\textrm{with}~i=j,\\
P_{\beta\alpha}& \textrm{if}~~ij=kl~\textrm{with}~i<j,\\
Q_{\beta\alpha}& \textrm{if}~~ij=kl~\textrm{with}~i>j,\\
pT_{\beta\alpha}& \textrm{if}~~ij=lk~\textrm{with}~i<j,\\
qT_{\beta\alpha}& \textrm{if}~~ij=lk~\textrm{with}~i>j,\\
0 & \textrm{for all other cases}.
\end{cases}
\end{equation*}
If $\mathbf{U}(X;t)=\mathbf{U}(x_1,\dots,x_N;t)$ be an $N^N \times N^N$ matrix whose entries are functions on $ \mathbb{Z}^N \times [0,\infty)$ which satisfies the differential equation
\begin{gather}
 \frac{{\rm d}}{{\rm d}t}\mathbf{U}(X;t) = \sum_{i=1}^N\big[p\mathbf{U}(x_1,\dots,x_{i-1},x_i-1,x_{i+1},\dots,x_N;t)\nonumber\\
\hphantom{\frac{{\rm d}}{{\rm d}t}\mathbf{U}(X;t) = \sum_{i=1}^N}{} + q\mathbf{U}(x_1,\dots,x_{i-1},x_i+1,x_{i+1},\dots,x_N;t)\big]
 - N \mathbf{U}(x_1,\dots,x_N;t),\label{454pm91}
\end{gather}
and the initial condition
\begin{equation}\label{634pm93}
\mathbf{U}(x_1,\dots,x_N;0) = \begin{cases}
\mathbf{I}_{N^N} & \textrm{if}~(x_1,\dots,x_N) = (y_1,\dots,y_N) ~\textrm{and}~x_1<\cdots<x_N, \\
\mathbf{0}_{N^N} & \textrm{if}~(x_1,\dots,x_N) \neq (y_1,\dots,y_N) ~\textrm{and}~x_1<\cdots<x_N
\end{cases}
\end{equation}
for a given initial positions of particles $(y_1,\dots,y_N)$ with $y_1<\cdots<y_N,$ and the \textit{boundary condition}
\begin{gather}
p\mathbf{U}(x_1,\dots,x_{i-1},x_i,x_i,x_{i+2},\dots,x_N;t)+q\mathbf{U}(x_1,\dots,x_{i-1},x_i+1,x_i+1,x_{i+2},\dots,x_N;t)\nonumber\\
\qquad{} = \big(\mathbf{I}_N^{\otimes{(i-1)}} \otimes \mathbf{B}\otimes \mathbf{I}_N ^{\otimes(N-i-1)}\big)\mathbf{U}(x_1,\dots,x_{i-1},x_i,x_i+1,x_{i+2},\dots,x_N;t)\label{253pm94}
\end{gather}
for all $i=1,{\dots}, N{-}1$, then we may assert that the restriction of $\mathbf{U}(x_1,{\dots},x_N;t)$ on $\{(x_1,{\dots}, x_N) \allowbreak \in \mathbb{Z}^N\colon x_1 < \cdots < x_N\}$ is $\mathbf{P}_Y(X;t)$. The solution of (\ref{454pm91}) by the Bethe ansatz is given
\begin{equation}
\sum_{\sigma\in {S}_N}\mathbf{A}_{\sigma}\prod_{i=1}^N\big(\xi_{\sigma(i)}^{x_i-y_{\sigma(i)}-1}{\rm e}^{\varepsilon(\xi_i) t}\big) , \label{455pm91}
\end{equation}
where
\begin{equation*}
\varepsilon(\xi_i) = \frac{p}{\xi_i} + q\xi_i -1
\end{equation*}
for $\xi_i \in \mathbb{C}\setminus \{0\}$, where $\mathbf{A}_{\sigma}$ is an $N^N \times N^N$ matrix whose entries are independent of $x_1,\dots,x_N$ and~$t$. Next, we will construct the matrix~$\mathbf{A}_{\sigma}$ so that~(\ref{455pm91}) satisfies the boundary condition~(\ref{253pm94}).
\begin{description}\itemsep=0pt
\item[Step 1.] Let $T_i$ be a simple transposition which interchanges the $i$th entry and the $(i+1)st$ entry and leaves the other entries fixed, that is,
\begin{equation*}
T_i\sigma = \sigma',
\end{equation*}
where $\sigma = \sigma(1)\cdots\sigma(i)\sigma(i{+}1){\cdots}\sigma(N)$ and $\sigma'= \sigma(1){\cdots}\sigma(i{-}1)\sigma(i{+}1)\sigma(i)\sigma(i{+}2){\cdots}\sigma(N)$. Since $T_1,\dots,T_{N-1}$ generate $S_N$, there is a finite sequence $a_1,\dots, a_n$ where each $a_i$ belongs to $\{1,2,\dots,N-1\}$ such that
\begin{equation*}
\sigma = T_{a_n}\cdots T_{a_1} %\label{1140-pm-523}
\end{equation*}
for any given $\sigma \in S_N$.
\item [Step 2.] For each $1\leq l\leq N-1$, we define
\begin{equation}
 \mathbf{T}_{l}({\beta},{\alpha}):= \mathbf{I}_N^{\otimes(l-1)} \otimes \mathbf{R}_{\beta\alpha} \otimes \mathbf{I}_N^{\otimes(N-l-1)}. \label{1243-am-519}
 \end{equation}
 \item [Step 3.] Let us denote $T_{a_k}\cdots T_{a_1}$ by $\sigma^{(k)}$ and let $\sigma^{(0)}$ be the identity permutation. Define
\begin{equation}\label{148am511}
\begin{aligned}
 \mathbf{A}_{\sigma}:= \mathbf{T}_{a_n}\big(\sigma^{(n-1)}(a_n+1),\sigma^{(n-1)}(a_n)\big) \cdots \mathbf{T}_{a_1}\big(\sigma^{(0)}(a_1+1),\sigma^{(0)}(a_1)\big).
\end{aligned}
\end{equation}
\end{description}

\begin{Lemma}
If $\mathbf{A}_{\sigma}$ is given as in \eqref{148am511}, then \eqref{455pm91} satisfies \eqref{253pm94} for each~$i$.
\end{Lemma}
\begin{proof}
Substituting (\ref{455pm91}) into (\ref{253pm94}) for $i$,
\begin{gather}
 \sum_{\sigma \in S_N}\big(p\mathbf{A}_{\sigma}\xi_{\sigma(1)}^{x_1}\cdots\xi_{\sigma(i-1)}^{x_{i-1}} \xi_{\sigma(i)}^{x_{i}}\xi_{\sigma(i+1)}^{x_{i}}\xi_{\sigma(i+2)}^{x_{i+2}}\cdots\xi_{\sigma(N)}^{x_{N}}\nonumber\\
 \qquad{} + q\mathbf{A}_{\sigma}\xi_{\sigma(1)}^{x_1}\cdots\xi_{\sigma(i-1)}^{x_{i-1}} \xi_{\sigma(i)}^{x_{i}+1}\xi_{\sigma(i+1)}^{x_{i}+1}\xi_{\sigma(i+2)}^{x_{i+2}}\cdots\xi_{\sigma(N)}^{x_{N}} \label{715pm94}\\
\qquad{} - \big(\mathbf{I}_N^{\otimes{(i-1)}} \otimes \mathbf{B}\otimes \mathbf{I}_N ^{\otimes(N-i-1)}\big)\mathbf{A}_{\sigma}\xi_{\sigma(1)}^{x_1}\cdots\xi_{\sigma(i-1)}^{x_{i-1}} \xi_{\sigma(i)}^{x_{i}}\xi_{\sigma(i+1)}^{x_{i}+1}\xi_{\sigma(i+2)}^{x_{i+2}}\cdots\xi_{\sigma(N)}^{x_{N}}\big) = \mathbf{0}_{N^N}.\nonumber
\end{gather}
If $\sigma'$ is an even permutation, then $T_i\sigma'$ is an odd permutation, and vice-versa. So, if we express~(\ref{715pm94}) as a sum over the alternating group~$A_N$,
\begin{gather}
 \sum_{\sigma' \in A_N}\big(\big[\mathbf{I}_{N^N}\big(p+q\xi_{\sigma'(i)}\xi_{\sigma'(i+1)}\big)
 - \big(\mathbf{I}_N^{\otimes{(i-1)}} \otimes \mathbf{B}\otimes \mathbf{I}_N ^{\otimes(N-i-1)}\big)\xi_{\sigma'(i+1)}\big]\mathbf{A}_{\sigma'}\nonumber \\
\qquad {} + \big[\mathbf{I}_{N^N}\big(p+q\xi_{\sigma'(i+1)}\xi_{\sigma'(i)}\big) - \big(\mathbf{I}_N^{\otimes{(i-1)}} \otimes \mathbf{B}\otimes \mathbf{I}_N ^{\otimes(N-i-1)}\big)\xi_{\sigma'(i)}\big]\mathbf{A}_{T_i\sigma'}\big)\nonumber\\
\qquad {} \times \xi_{\sigma'(1)}^{x_1}\cdots\xi_{\sigma'(i-1)}^{x_{i-1}}\xi_{\sigma'(i)}^{x_{i}} \xi_{\sigma'(i+1)}^{x_{i}}\xi_{\sigma'(i+2)}^{x_{i+2}}\cdots \xi_{\sigma'(N)}^{x_{N}} = \mathbf{0}_{N^N}.\label{1117pm95}
\end{gather}
A sufficient condition for (\ref{1117pm95}) is that for each $\sigma \in S_N$,
\begin{gather}
\mathbf{A}_{T_i\sigma} = - \big[\mathbf{I}_{N^N}\big(p+q\xi_{\sigma(i+1)}\xi_{\sigma(i)}\big) - \big(\mathbf{I}_N^{\otimes{(i-1)}} \otimes \mathbf{B}\otimes \mathbf{I}_N ^{\otimes(N-i-1)}\big)\xi_{\sigma(i)}\big]^{-1}\nonumber\\
\hphantom{\mathbf{A}_{T_i\sigma} =}{} \times \big[\mathbf{I}_{N^N}\big(p+q\xi_{\sigma(i)}\xi_{\sigma(i+1)}\big) - \big(\mathbf{I}_N^{\otimes{(i-1)}} \otimes \mathbf{B}\otimes \mathbf{I}_N ^{\otimes(N-i-1)}\big)\xi_{\sigma(i+1)}\big] \mathbf{A}_{\sigma}\nonumber \\
\hphantom{\mathbf{A}_{T_i\sigma}}{} = -\big(\mathbf{I}_{N^{(i-1)}}\otimes \big[\mathbf{I}_{N^2}\big(p+q\xi_{\sigma(i+1)}\xi_{\sigma(i)}\big) - \mathbf{B} \xi_{\sigma(i)} \big]^{-1} \otimes \mathbf{I}_{N^{(N-i-1)}}\big)\nonumber \\
\hphantom{\mathbf{A}_{T_i\sigma} =}{} \times \big(\mathbf{I}_{N^{(i-1)}}\otimes \big[\mathbf{I}_{N^2}\big(p+q\xi_{\sigma(i+1)}\xi_{\sigma(i)}\big) - \mathbf{B} \xi_{\sigma(i+1)} \big] \otimes \mathbf{I}_{N^{(N-i-1)}}\big)\mathbf{A}_{\sigma} \nonumber\\
\hphantom{\mathbf{A}_{T_i\sigma} }{}= \big(\mathbf{I}_{N^{(i-1)}}\otimes \mathbf{R}_{\sigma(i+1)\sigma(i)} \otimes \mathbf{I}_{N^{(N-i-1)}}\big)\mathbf{A}_{\sigma}. \label{1151pm95}
\end{gather}
Now, it remains to show that if $\mathbf{A}_{\sigma}$ is given by (\ref{148am511}), then (\ref{1151pm95}) is satisfied. This is immediately obtained because
\begin{gather*}
\mathbf{A}_{T_i\sigma} = \mathbf{T}_i\mathbf{T}_{a_n} \cdots \mathbf{T}_{a_1} = \mathbf{T}_i\big(\sigma(i+1),\sigma(i)\big)\mathbf{A}_{\sigma} \\
\hphantom{\mathbf{A}_{T_i\sigma}}{} = \big(\mathbf{I}_{N^{(i-1)}}\otimes \mathbf{R}_{\sigma(i+1)\sigma(i)} \otimes \mathbf{I}_{N^{(N-i-1)}}\big)\mathbf{A}_{\sigma}
\end{gather*}
by (\ref{1243-am-519}).
\end{proof}
\begin{Remark}
The consistency relations
\begin{gather}
 \mathbf{T}_i(\beta,\alpha)\mathbf{T}_j(\delta,\gamma) = \mathbf{T}_j(\delta,\gamma)\mathbf{T}_i(\beta,\alpha) \qquad \textrm{if}~|i - j| \geq 2,\nonumber \\
\mathbf{T}_i(\gamma,\beta)\mathbf{T}_j(\gamma,\alpha)\mathbf{T}_i(\beta,\alpha) = \mathbf{T}_j(\beta,\alpha) \mathbf{T}_i(\gamma,\alpha)\mathbf{T}_j(\gamma,\beta) \qquad \textrm{if}~|i-j| = 1, \nonumber\\
\mathbf{T}_i(\beta,\alpha) \mathbf{T}_i(\alpha,\beta) = \mathbf{I}_{N^N}\label{617pm93}
\end{gather}
can be directly verified by the matrix multiplication. In particular, (\ref{617pm93}) with $p=1$ was verified in~\cite{Chatterjee-Schutz-2010}.
\end{Remark}

Now, we choose $\mathbf{A}_{1\cdots N}= \mathbf{I}_{N^N}$ and take a multiple-fold contour integral of each element of the
matrix (\ref{148am511}) with respect to $\xi_1,\dots, \xi_N$ over sufficiently small circles centered at the origin, denoting the matrix by
\begin{equation}\label{1223-am-519}
 \dashint_c\cdots \dashint_c\sum_{\sigma\in {S}_N}\mathbf{A}_{\sigma}\prod_{i=1}^N\big(\xi_{\sigma(i)}^{x_i-y_{\sigma(i)}-1}{\rm e}^{\varepsilon(\xi_i) t}\big){\rm d}\xi_1\cdots {\rm d}\xi_N.
\end{equation}
Finally, it should be proved that (\ref{1223-am-519}) satisfies the initial condition (\ref{634pm93}), but its proof is essentially the same as the proof for the ASEP in \cite{Tracy-Widom-2011,Tracy-Widom-2013} because the poles arising in the integrands are the same as the poles in the ASEP formulas. So, we omit the proof and conclude that
\begin{equation}\label{1223-am-51911}
 \mathbf{P}_Y(X;t) = \dashint_c\cdots \dashint_c\sum_{\sigma\in {S}_N}\mathbf{A}_{\sigma}\prod_{i=1}^N\big(\xi_{\sigma(i)}^{x_i-y_{\sigma(i)}-1}{\rm e}^{\varepsilon(\xi_i) t}\big) {\rm d}\xi_1\cdots {\rm d}\xi_N.
\end{equation}

\section{Determinantal formulas}\label{det}
In this section we focus on the diagonal of the matrix (\ref{1223-am-51911}) for the totally asymmetric case $p=1$, that is, the transition probabilities $P_{(Y,\pi)}(X,\pi;t)$. In the case of the totally asymmetric model ($p=1$), if the order of particles at $t=0$ is represented by a permutation $\pi$ of a given multi-set and the order of particles at time $t=t_0>0$ is also given by $\pi$, then the order of particles at each time $0\leq t\leq t_0$ must be $\pi$. If all particles belong to the same species, then $P_{(Y,i\cdots i)}(X,i\cdots i;t)$ is simply the TASEP's transition probability. Sch\"{u}tz found its determinantal formula~\cite{Schutz-1997}. For the TASEP with second class particles (two-species TASEP), Chatterjee and Sch\"{u}tz found the determinantal formulas for the transition probabilities when there is no change in order of particles \cite{Chatterjee-Schutz-2010}. We will extend Chatterjee and Sch\"{u}tz's result to the multi-species TASEP. Let
\begin{equation*}
F_{n}(x;t) = \dashint_c \xi^{x-1}(1-\xi)^{-n}{\rm e}^{(1/\xi - 1)t}{\rm d}\xi.
\end{equation*}
For a given permutation $\pi=\pi_1\cdots \pi_N$ of a multi-set, let $\#_{ij}$ be the number of pairs of places $(i,i+1)$ between the $i$th place and $j$th place such that $\pi_i > \pi_{i+1}$. For example, if $\pi = 214311$, then $\#_{12} = 1$, $\#_{14} = 2,$ and $\#_{15} = \#_{16} = 3$. Obviously, $\#_{ij}= \#_{ji}$. Let
\begin{equation*}
n(i,j)=\operatorname{sgn}(i - j) \times (|i - j| -\#_{ij}).
\end{equation*}
If we denote the diagonal elements of $\mathbf{U}(x_1,\dots, x_N;t)$ by $U_{\pi,\pi}(x_1,\dots,x_N;t)$ where $\pi=\pi_1\cdots \pi_N$, then the equations for the diagonal of (\ref{454pm91}), (\ref{634pm93}) and (\ref{253pm94}) are
\begin{gather}
 \frac{{\rm d}}{{\rm d}t}U_{\pi,\pi}(x_1,\dots,x_N;t) = \sum_{i=1}^NU_{\pi,\pi}(x_1,\dots,x_{i-1},x_i-1,x_{i+1},\dots,x_N;t)\nonumber\\
 \hphantom{\frac{{\rm d}}{{\rm d}t}U_{\pi,\pi}(x_1,\dots,x_N;t) =}{} - NU_{\pi,\pi}(x_1,\dots,x_N;t),\label{503am814}
\\
\label{209pm98}
U_{\pi,\pi}(x_1,\dots, x_N;0) =\prod_{i=1}^N\delta_{x_iy_i} \qquad \textrm{when}\quad x_1<\cdots<x_N~~\textrm{for given $y_1<\cdots < y_N$}
\end{gather}
and
\begin{gather}
U_{\pi,\pi}(x_1,\dots, x_i,x_i,x_{i+2},\dots,x_N;t) \nonumber\\
\qquad{} = \begin{cases}
 U_{\pi,\pi}(x_1,\dots,{x_i},x_{i}+1,x_{i+2},\dots,x_N;t), & \textrm{if $\pi_i \leq \pi_{i+1}$}, \\
 0& \textrm{if $\pi_i> \pi_{i+1}$}.\label{548am814}
\end{cases}
\end{gather}

The following theorem gives the solution to (\ref{503am814}), (\ref{209pm98}), and (\ref{548am814}).
\begin{Theorem}
For given $Y = (y_1,\dots, y_N)$ with $y_1<\cdots<y_N$ and $\pi$, define
\begin{gather*}
G(x_1,\dots, x_N;t) = \left[
 \begin{matrix}
 F_{n(1,1)}(x_1 - y_1;t)\!\! & F_{n(1,2)}(x_1 - y_2;t)\!\! & \!\!\cdots\!\! & F_{n(1,N)}(x_1 - y_N;t) \\
 F_{n(2,1)}(x_2 - y_1;t)\!\! & F_{n(2,2)}(x_2 - y_2;t)\!\! & \!\!\cdots\!\! & F_{n(2,N)}(x_2 - y_N;t) \\
 \vdots & \vdots & \!\!\ddots\!\! & \vdots \\
 F_{n(N,1)}(x_N - y_1;t)\!\! & F_{n(N,2)}(x_N - y_2;t)\!\! & \!\!\cdots\!\! & F_{(N,N)}(x_N - y_N;t)
 \end{matrix}
 \right].
\end{gather*}
Then,
\begin{equation*}
P_{(Y,\pi)}(x_1,\dots,x_N,\pi;t) = \det G(x_1,\dots, x_N;t).
\end{equation*}
\end{Theorem}
\begin{proof}
(i) To show that $\det G(x_1,\dots, x_N;t)$ satisfies (\ref{503am814}).
Since
\begin{equation*}
\frac{{\rm d}}{{\rm d}t}F_{n(i,j)}(x_i - y_j;t) = -F_{n(i,j)}(x_i - y_j;t) + F_{n(i,j)}(x_i -1 - y_j;t),
\end{equation*}
we have
\begin{gather*}
\frac{{\rm d}}{{\rm d}t} \det G(x_1,\dots, x_N;t) \\
\qquad{} = \left|
 \begin{matrix}
\dfrac{{\rm d}}{{\rm d}t} F_{n(1,1)}(x_1 - y_1;t) & \dfrac{{\rm d}}{{\rm d}t}F_{n(1,2)}(x_1 - y_2;t) & \cdots & \dfrac{{\rm d}}{{\rm d}t} F_{n(1,N)}(x_1 - y_N;t) \vspace{1mm}\\
 F_{n(2,1)}(x_2 - y_1;t) & F_{n(2,2)}(x_2 - y_2;t) & \cdots & F_{n(2,N)}(x_2 - y_N;t) \\
 \vdots & \vdots & \ddots & \vdots \\
 F_{n(N,1)}(x_N - y_1;t) & F_{n(N,2)}(x_N - y_2;t) & \cdots & F_{(N,N)}(x_N - y_N;t)
 \end{matrix}
 \right| +\cdots \\
\qquad\quad{} + \left|
 \begin{matrix}
 F_{n(1,1)}(x_1 - y_1;t) &F_{n(1,2)}(x_1 - y_2;t) & \cdots & F_{n(1,N)}(x_1 - y_N;t) \\
F_{n(2,1)}(x_2 - y_1;t) & F_{n(2,2)}(x_2 - y_2;t) & \cdots & F_{n(2,N)}(x_2 - y_N;t) \\
 \vdots & \vdots & \ddots & \vdots \\
\dfrac{{\rm d}}{{\rm d}t} F_{n(N,1)}(x_N - y_1;t) & \dfrac{{\rm d}}{{\rm d}t} F_{n(N,2)}(x_N - y_2;t) & \cdots & \dfrac{{\rm d}}{{\rm d}t} F_{(N,N)}(x_N - y_N;t)
 \end{matrix}
 \right| \\
 \qquad{} = -N \det G(x_1,\dots, x_N;t) +\sum_{i=1}^N\det G(x_1,\dots,x_{i-1},x_i-1,x_{i+1},\dots, x_N;t).
\end{gather*}

(ii) To show that $\det G(x_1,\dots, x_N;t)$ satisfies (\ref{548am814}): Suppose that $\pi_i>\pi_{i+1}$. First, we will show that $n(i,j) = n(i+1,j)$. If $j>i+1$, then
\begin{equation*}
n(i,j) = (-1) \times (j-i -\#_{ij}) = (-1) \times \big[j-(i+1) -(\underbrace{\#_{ij}-1}_{\#_{(i+1)j}})\big] = n(i+1,j).
\end{equation*}
If $j<i-1$, then
\begin{equation*}
n(i,j) = 1 \times (i-j -\#_{ij}) = 1 \times \big[i+1-j -(\underbrace{\#_{ij}+1}_{\#_{(i+1)j}})\big] = n(i+1,j).
\end{equation*}
If $i=j$ or $j=i\pm 1$, then
\begin{equation*}
n(i,j) = 0 = n(i+1,j).
\end{equation*}
Hence, the $i$th row and the $(i+1)$th row of $G(x_1,\dots, x_i,x_i,x_{i+2},\dots,x_N;t)$ are the same and
\begin{equation*}
\det G(x_1,\dots, x_i,x_i,x_{i+2},\dots,x_N;t) = 0.
\end{equation*}
 Next, suppose that $\pi_i \leq \pi_{i+1}$, so $\#_{ij} = \#_{(i+1)j}$. We will show that $n(i,j) = n(i+1,j)-1$.
If $j>i+1$, then
\begin{equation*}
n(i+1,j) = (-1) \times (j-i-1 -\#_{(i+1)j}) = (-1) \times (j-i -\#_{ij}) + 1 = n(i,j)+1.
\end{equation*}
If $j<i-1$, then
\begin{equation*}
n(i+1,j) = 1 \times (i+1-j -\#_{(i+1)j}) = 1 \times (i-j -\#_{ij}) +1 = n(i,j) +1.
\end{equation*}
If $i=j$, then $n(i+1,j) = 1 = n(i,i) +1$. If $j = i +1$, then $n(i+1,j) = 0 = n(i,j) +1.$ If $j = i-1$, then
\begin{gather}
n(i+1,j) = (+1)\times (2 - \#_{(i-1)(i+1)})= (+1)\times (2 - \#_{(i-1)i}) \nonumber\\
 \hphantom{n(i+1,j)}{} =(+1)\times (1 - \#_{(i-1)i}) + 1 = n(i,i-1) + 1 = n(i,j)+1.\label{740pm815}
\end{gather}
Also, using the identity
\begin{equation*}
F_n(x+1;t) = F_n(x;t) - F_{n-1}(x;t),
\end{equation*}
we obtain
\begin{gather}
 \left|
 \begin{matrix}
 \vdots & \vdots & \vdots \\
 F_{n(i,1)}(x_i - y_1;t) & \cdots & F_{n(i,N)}(x_i - y_N;t) \\
 F_{n(i+1,1)}(x_{i}+1 - y_1;t) & \cdots & F_{n(i+1,N)}(x_{i}+1 - y_N;t) \\
 \vdots & \vdots & \vdots
 \end{matrix}
 \right| \nonumber\\
\qquad{} = \left|
 \begin{matrix}
 \vdots & \vdots & \vdots \\
 F_{n(i,1)}(x_i - y_1;t) & \cdots & F_{n(i,N)}(x_i - y_N;t) \\
 F_{n(i+1,1)}(x_{i} - y_1;t) & \cdots & F_{n(i+1,N)}(x_{i} - y_N;t) \\
 \vdots & \vdots & \vdots \\
 \end{matrix}
 \right| \nonumber\\
\qquad\quad{} - \left|
 \begin{matrix}
 \vdots & \vdots & \vdots \\
 F_{n(i,1)}(x_i - y_1;t) & \cdots & F_{n(i,N)}(x_i - y_N;t) \\
 F_{n(i+1,1)-1}(x_{i} - y_1;t) & \cdots & F_{n(i+1,N)-1}(x_{i} - y_N;t) \\
 \vdots & \vdots & \vdots
 \end{matrix}
 \right|, \label{745pm815}
\end{gather}
where the second term on the right-hand side of (\ref{745pm815}) is zero by (\ref{740pm815}). Hence,
\begin{equation*}
\det G(x_1,\dots, x_i,x_i+1,x_{i+2},\dots,x_N;t) = \det G(x_1,\dots, x_i,x_i,x_{i+2},\dots,x_N;t).
\end{equation*}

(iii) To show that $\det G(x_1,\dots, x_N)$ satisfies (\ref{209pm98}): Since $F_{n(i,i)}(x_i - y_i;0) = \delta_{x_iy_i}$ and $F_{n(i,j)}(x_i - y_j;0) =0$ for $i>j$ because $x_i - y_j \geq 1$ when $x_1<\cdots <x_N$ for given $y_1<\cdots <y_N$,
 \begin{equation*}
 \det G(x_1,\dots,x_N;0) = \prod_{i=1}^N\delta_{x_iy_i}
 \end{equation*}
 when $x_1<\cdots <x_N$ for given $y_1<\cdots <y_N$.
\end{proof}

When $\pi = (1~1~\cdots~1)$, that is, for the TASEP with single species, $F_{n(i,j)}(x_i - y_j;t) = F_{i-j}(x_i - y_j;t)$ has some interesting properties, for example, \cite[equation~(4)]{Sasamoto-2004}, which are needed to find one-point distributions and multi-point distributions \cite{Sasamoto-2004,Rakos-Schutz-2006, Sasamoto-2005}. At this moment, it is not clear if $F_{n(i,j)}(x_i - y_j;t)$ has similar properties for general $\pi$. It would be interesting to see if there is a $\pi$ other than $(1~1~\cdots~ 1)$ whose $F_{n(i,j)}(x_i - y_j;t)$ has nice properties to enable us to find the one-point and the multi-point distributions.
Also, it is notable that Kuan recently obtained a determinantal expression for a multi-point distribution in a different context (that is, of \textit{the maximum species number among all particles to the left of a point} (see \cite{Kuan-2020} for detailed definition) in the inhomogeneous multi-species TASEP \cite{Kuan-2020}. But, the determinantal formulas in this paper are for the transition probabilities ($\pi \rightarrow \pi$) for all possible cases that particles belong to some species.

\subsection*{Acknowledgements}

This work was supported by the faculty development competitive research grant (090118FD5341) by Nazarbayev University. The author is grateful for valuable comments from Jeffrey Kuan and Atsuo Kuniba on the multi-species ASEP. This research was supported in part by the International Centre for Theoretical Sciences (ICTS) at Bengaluru, India during a visit for participating in the program~-- Universality in random structures: Interfaces, Matrices, Sandpiles in 2019.

\pdfbookmark[1]{References}{ref}
\LastPageEnding

\end{document}